\documentclass[smallextended]{svjour3}      

\smartqed 

\usepackage{graphicx}

\journalname{Foundations of Physics}

\begin{document}

\title{The matter-gravity entanglement hypothesis}

\author{Bernard S. Kay}

\institute{\at Department of Mathematics, University of York, York YO10 5DD, UK \\
\email{bernard.kay@york.ac.uk} }

\date{Received: date / Accepted: date}

\maketitle

\begin{abstract} 
I outline some of my work and results (some dating back to 1998, some more recent) on my matter-gravity entanglement hypothesis, according to which the entropy of a closed quantum gravitational system is equal to the system's matter-gravity entanglement entropy.  The main arguments presented are:  (1) that this hypothesis is capable of resolving what I call the second-law puzzle, i.e.\ the puzzle as to how the entropy increase of a closed system can be reconciled with the asssumption of unitary time-evolution;   (2) that the black hole information loss puzzle may be regarded as a special case of this second law puzzle and that therefore the same resolution applies to it;  (3) that the black hole thermal atmosphere puzzle (which I recall) can be resolved by adopting a radically different-from-usual description of quantum black hole equilibrium states, according to which they are total pure states, entangled between matter and gravity in such a way that the partial states of matter and gravity are each approximately thermal equilibrium states (at the Hawking temperature);   (4) that the Susskind-Horowitz-Polchinski string-theoretic understanding of black hole entropy as the logarithm of the degeneracy of a long string (which is the weak string coupling limit of a black hole) cannot be quite correct but should be replaced by a modified understanding according to which it is the entanglement entropy between a long string and its stringy atmosphere, when in a total pure equilibrium state in a suitable box, which (in line with (3)) goes over, at strong-coupling, to a black hole in equilibrium with its thermal atmosphere.  The modified understanding in (4) is based on a general result, which I also describe, which concerns the likely state of a quantum system when it is weakly coupled to an energy-bath and the total state is a random pure state with a given energy.   This result generalizes Goldstein et al.'s `canonical typicality' result to systems which are not necessarily small.
\end{abstract}

\section{The second law puzzle}

Let me begin my talk\footnote{This article is a written version of a talk given at the 18th UK and European Conference on Foundations of Physics (16-18 July 2016, LSE, London)} by recalling one version of the second law of thermodynamics:  

\medskip

\noindent
\textit{The entropy of the universe begins low and increases monotonically.}

\medskip

There are long-established and well-known arguments -- see the discussion of `branch systems' in \cite{Reichenbach} as also reviewed e.g.\ in \cite{Davies}) -- that other statements of the second law, in terms of what can and can't happen with heat engines, refrigerators etc.\ follow from the above statement.   As also explained in these references, the above statement leads to an explanation of time asymmetry; i.e.\ why, for example, it is commonplace to observe wine-glasses fall off tables and smash into pieces, but we never see lots of smashed pieces assemble themselves into wine-glasses and jump onto tables.   

\begin{figure}[h]
\centering
\includegraphics[trim = 7cm 20cm 5cm 5cm, clip]{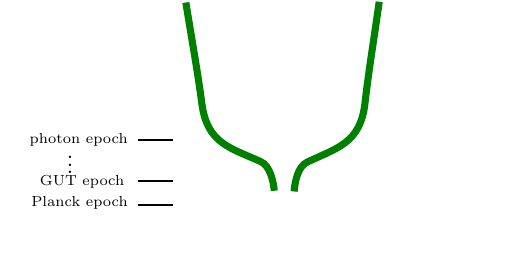}
\caption{Schematic diagram of the universe showing how its radius increases with time.}
\label{fig:1}      
\end{figure}

But how do we define the entropy of a closed system? And why \textit{does it} increase?

\begin{figure}[h]
\centering
\includegraphics[trim = 6cm -0.5cm 6cm -1cm, clip]{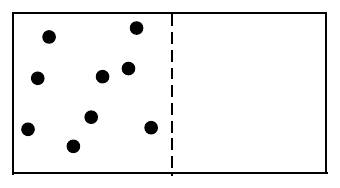}
\caption{A box of gas molecules, initially confined to the left half.}
\label{fig:2}      
\end{figure}

A standard way of answering this (essentially due to Boltzmann around 1870) might be to consider for example what will happen if one starts with a system of $N$ gas molecules in the left half of a box (see Figure \ref{fig:2}) and removes a partition, allowing the particles to diffuse into the right half of the box.

\begin{figure}[h]
\centering
\includegraphics[scale=1.4, trim = 6cm 22.3cm 6cm 4cm, clip]{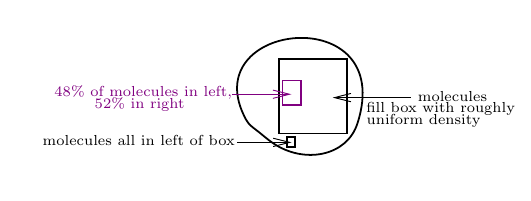}
\caption{The phase space for the gas in the box, indicating some possible macrostates.}
\label{fig:3}      
\end{figure}

In a classical discussion, one describes the states of this system with some given energy in terms of a 
$6N-1$ dimensional phase space, the points of which are called `microstates' and (see Figure \ref{fig:3}) one imagines this phase space to be divided up into cells -- called `macrostates' -- with the property that we cannot in practice distinguish between any pair of microstates in any single macrostate. One then defines the (`coarse-grained') entropy, $S$,  of a microstate by
\begin{equation}
\label{Boltzmann}
S = k\log W
\end{equation}
where $k$ is Boltzmann's constant and $W$ is the volume of the macrostate containing that given microstate.  

The standard argument then is that (see Figure \ref{fig:3}) the macrostate corresponding to ``all the particles are in the left half of the box'' will have a vastly smaller volume in phase space than the large macrostate which corresponds to ``the molecules fill the box with roughly uniform density''.   Hence, as time goes on and the state of the system wanders around the phase space accordingly, it is highly likely that the entropy -- as defined by (\ref{Boltzmann}) will get bigger and stay bigger.

However, this definition of entropy and this argument for its increase depends, unsatisfactorily, on the need to make judgments about what \textit{we} can distinguish.   For example, if (see again Figure \ref{fig:3}) after previously ignoring such fine distinctions, we were to take the view that we can distinguish a state where, say, 48\% of the particles are in the left half of the box and 52\% in the right half from a state with roughly equal proportions\footnote{These numbers were not entirely randomly chosen, the talk being given shortly after the June 2016 Brexit referendum.} then, at times for which the system's microstate lies in the accordingly-defined new macrostate (obviously a subregion of the previously discussed large macrostate) then equation (\ref{Boltzmann}) would ascribe a different value to the entropy.

Moreover, this unsatisfactory arbitrariness and vagueness in the definition of entropy is even more of a problem if we want to account for the version of the second law with which we began.  For \textit{we} are not even present to make any distinctions in the early universe! 

Turning to the quantum setting,  von Neumann gave us long ago a quantum translation of Boltzmann's equation (\ref{Boltzmann}).   Given a description of our system in terms of a density operator, $\rho$ acting on the system's Hilbert space ${\cal H}$, one defines its von Neumann entropy, $S^{\mathrm{vN}}(\rho)$, by 
\begin{equation}
\label{vN}
S^{\mathrm{vN}}(\rho)=-k{\rm tr}(\rho\log\rho).
\end{equation}
But if we were to equate the physical entropy, $S^{\mathrm{physical}}$, with $S^{\mathrm{vN}}(\rho)$ and if $\rho$ satisfies the usual unitary time evolution rule
\[
\rho(t)=U(t)\rho(0)U(t)^{-1}
\]
then we would conclude that
\[
S^{\mathrm{physical}}(\rho(t))= {\mathrm{constant}}.
\]
in contradiction with the second law.   We shall call this the \textit{second law puzzle}.
One can overcome this difficulty by defining quantum counterparts to the above classical coarse-graining, but of course one then would have the same unsatisfactory vagueness and subjectivity as we discussed above in the classical case.   

More interestingly, one can seek to exploit a feature of quantum mechanics which has no classical counterpart:   If we have a pure state, described by a density operator, $\rho=|\Psi\rangle\langle\Psi|$, which is a projector onto a vector, $\Psi$, in a Hilbert space, ${\cal H}_{\mathrm{total}}$, which arises as the tensor product, 
\[
{\cal H}_{\mathrm{total}}={\cal H}_\mathrm{A}\otimes{\cal H}_\mathrm{B}
\]
 of two Hilbert spaces, ${\cal H}_\mathrm{A}$ and ${\cal H}_\mathrm{B}$,  then the reduced density operator, $\rho_A$ on ${\cal H}_\mathrm{A}$, defined as the partial trace, ${\rm tr}_{{\cal H}_\mathrm{B}}(\rho)$, of $\rho$ over ${\cal H}_\mathrm{B}$, will typically have $S^{\mathrm{vN}}(\rho_\mathrm{A}) > 0$.

We remark that

\begin{itemize}

\item {This partial trace is characterized by the property that, if $O$ is a (self-adjoint) operator on ${\cal H}_\mathrm{A}$, then  
\[
{\rm tr}(\rho_\mathrm{A} O)_{{\cal H}_\mathrm{A}}= \langle\Psi(O\otimes I)|\Psi\rangle_{{\cal H}_{\mathrm{total}}}.
\]}

\smallskip

\item{Both reduced density operators have equal von Neumann entropies:
\begin{equation}
\label{equalents}
S^{\mathrm{vN}}(\rho_A)=S^{\mathrm{vN}}(\rho_\mathrm{B})
\end{equation}
and this common value is often known as the A-B entanglement entropy of the total state-vector $\Psi$.}

\end{itemize}

In a variant of the `environment paradigm for decoherence' or, from another point of view, a variant of a possible approach  to quantum statistical mechanics, this formalism is often applied in the case that A is interpreted as standing for some `system' and B for the system's  `environment' or `energy bath' and $S^{\mathrm{vN}}(\rho_\mathrm{A})$ is then interpreted as the entropy of the system due to entanglement with the environment.

So, for a fixed split of our total system into `system' and `environment' (i.e.\ a fixed expression of our total Hilbert space, $\cal H$ as a tensor product ${\cal H}_\mathrm{A}\otimes{\cal H}_\mathrm{B}$) the environment paradigm gives us an objective notion of entropy.  However, there remain problems:

\begin{itemize}

\item 	It only offers a notion of entropy for \textit{open} systems.

\smallskip

\item There are lots of ways of expressing a given Hilbert space, $\cal H$, as a tensor product ${\cal H}_\mathrm{A}\otimes{\cal H}_\mathrm{B}$.   How we choose a way -- i.e.\ how we split our total system into `system' and `environment' -- depends on subjective choices and, again,  \textit{we} are not around in the early universe to make those choices.

\end{itemize}

What I'd like to point out is that one can envisage an alternative physical use of this mathematical fact:  Suppose there's some decomposition that's physically natural, then maybe we could define the entropy of a total \textit{closed} system by
\begin{equation}
\label{3entropies}
 S^\mathrm{total}= S^{\mathrm{vN}}(\rho_\mathrm{A}) \quad (= S^{\mathrm{vN}}(\rho_\mathrm{B})) \quad  \hbox{(= A-B entanglement entropy)}
\end{equation}
\textit{rather than interpreting this mathematical quantity as the entropy of the A-subsystem}!

We propose that the identification:  
\[
\hbox{A=\textit{matter}; \quad B=\textit{gravity}},
\]
is the right choice.  This is our \textit{matter-gravity entanglement hypothesis}.  (See \cite{Kay1998a}, \cite{Kay1998b} and \cite{KayAbyaneh} for early papers, and \cite{KayEntropy} and the remainder of the present article for recent partial overviews and further references.)

In support of this, we note that the decomposition has to be meaningful throughout the entire history of the universe:  E.g.\ we couldn't identify A with \textit{photons} and  B with \textit{nuclei+electrons} because these notions are not even meaningful until the photon epoch.   We content ourselves, though, with going back to just after the Planck epoch; we assume that a low-energy quantum gravity theory holds there and throughout the entire subsequent history of the universe and that this is a conventional (unitary) quantum theory with ${\cal H}={\cal H}_{matter}\otimes{\cal H}_{gravity}$.    We will also assume that the initial degree of matter-gravity entanglement is low.  (We leave it for a future theory of the pre-Planck era to explain that.)  

These assumptions then appear to be capable of offering an explanation of the second law in the form stated at the outset since one can argue that an initial state with a low degree of matter-gravity entanglement will, because of matter-gravity interaction, get more entangled, plausibly monotonically, as time increases.   At least the question of whether the second law holds becomes a question which, in principle, can be answered mathematically once we specify the (low-energy) quantum gravity Hamiltonian (i.e.\ the generator of the unitary time-evolution) and the initial state.    What we have called the second law puzzle would then be resolved because once we define entropy as matter-gravity entanglement entropy (rather than as the von Neumann entropy of the total state) there is no conflict between its increase and a unitary time-evolution.

\section{The information loss puzzle (Hawking 1976)}

The celebrated result of Hawking \cite{HawkingCMP75} is that a black hole formed by the dynamical collapse of a star will emit thermal radiation at the Hawking temperature, given, in the case of a spherically symmetric electrically neutral black hole by

\begin{figure}[h]
\centering
\includegraphics[trim = 6cm 21cm 6cm 4cm, clip]{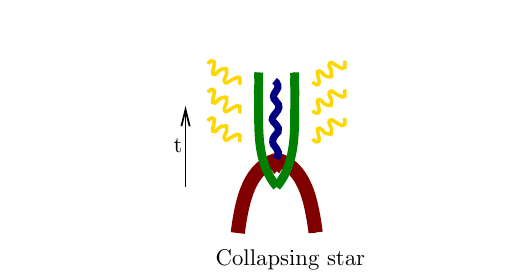}
\caption{A schematic picture of the spacetime of a star which collapses to a black hole and
then Hawking-evaporates.   The thick brown lines represent the boundary of the surface of a collapsing star, the green lines the horizon, the blue wiggly line the future spacetime singularity.  The thin yellow  wiggles indicate the Hawking radiation predicted in \cite{HawkingCMP75}.}
\label{fig:4}      
\end{figure}

\begin{equation}
\label{HawkTemp}
kT_\mathrm{Hawking}={1\over{8\pi GM}}
\end{equation}

where $M$ is the black hole mass (and we take $c=\hbar=1$).

As Hawking explained in that work, one expects that such a radiating black hole will lose mass, increasing further its temperature, and eventually evaporate.

During this whole process of collapse to a black hole and subsequent evaporation, one expects the entropy of the total system to increase monotonically.\footnote{Without wishing to imply that they are necessarily exactly additive, we note that while the entropy of the black hole (given by (\ref{HawkEnt})) will decrease because the horizon area will decrease, one expects that this will be more than compensated by the increased entropy of the sphere of emitted Hawking radiation which is growing in size at the speed of light and within which, moreover, the later radiation will be hotter than that emitted earlier.}

The version of the  \textit{information loss puzzle} \cite{HawkingInfoLoss} that I shall adopt here is the puzzle as to how this entropy increase can be reconciled with an assumption of unitary time evolution. 

Stated in this way, I think it is clear that the information loss puzzle is nothing but a special case of our Second Law Puzzle; we recall here that this is the puzzle that, if one equates $S^\mathrm{physical}$ with $S^{\mathrm{vN}}(\rho_\mathrm{total}$), then $S^\mathrm{physical}$ must be constant.

I suggested in \cite{Kay1998a} and \cite{Kay1998b} that the resolution to the information loss puzzle is simply a special case of the above proposed resolution to the second law puzzle.   Namely, $S^\mathrm{physical}$ is \textit{not} $S^{\mathrm{vN}}(\rho_\mathrm{total})$.   Rather $S^\mathrm{physical}$ is the total state's matter-gravity entanglement entropy.   As I already said in the more general context in Section 1, this is not a unitary invariant and -- it is reasonable to assume -- would increase.   Thus with our definition of entropy as matter-gravity entanglement entropy, we reconcile unitarity with information loss (i.e.\ with entropy increase) and there is no longer a puzzle.  That it also offers this resolution to the information loss puzzle, is, in my view, further evidence that our matter-gravity entanglement hypothesis is on the right track.

\section{The thermal atmosphere puzzle}

\begin{figure}[h]
\centering
\includegraphics[scale = 0.7, trim = 6cm 19cm 6cm 5cm, clip]{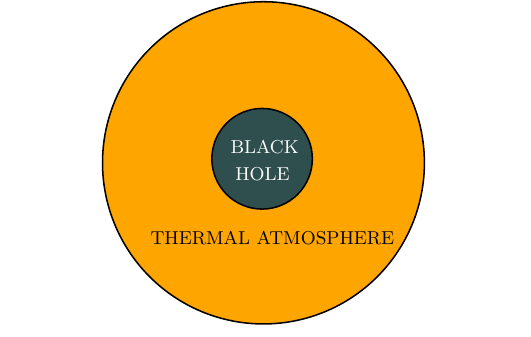}
\caption{A schematic picture of a black hole in equilibrium with its thermal atmosphere in a box.}
\label{fig:5}      
\end{figure}

A black hole in a box in equilibrium with its thermal atmosphere (see Figure \ref{fig:5}) is traditionally taken to be in a total Gibbs state (in particular a total mixed state) at the Hawking temperature

Everyone agrees that the entropy of this system has (at least up to small corrections) the value 
\begin{equation}
\label{HawkEnt}
S^\mathrm{Hawking}=4\pi kGM^2 = kA/4G.
\end{equation}
where $A$ is the surface area of the event horizon ($=16\pi G^2M^2$).
The thermal atmopshere puzzle \cite{Page} \cite{WaldLivRev} is that one can give seemingly convincing arguments for each of the following three, at first sight seemingly mutually contradictory, statements about the nature and origin of this entropy: 

\begin{itemize}

\item{\textit{It is the entropy of the gravitational field (so mostly `residing' in the black hole).}} 

\smallskip

\item{\textit{It is the entropy of the thermal atmosphere (so apart from the graviton component, consisting mainly of matter).}}

\smallskip
 
\item{\textit{It is the sum of the above two entropies.}}

\end{itemize}

Our proposed resolution of the puzzle begins by postulating that it is not actually the case that the total state is a Gibbs state; rather, we propose, \textit{the total state is pure}, but entangled between gravity ($\simeq$ the black hole) and matter ($\simeq$ its atmosphere) in such a way that each are approximately Gibbs states (at the Hawking temperature).   

We further suggest, in line with our matter-gravity entanglement hypothesis,  that $S^\mathrm{Hawking}$ is really this state's \textit{matter-gravity entanglement entropy}.   This offers to resolve the puzzle in the following way:  The first entropy can be regarded, according to the environment paradigm, as the entropy of the open system consisting of the gravitational field due to its matter environment; the second the entropy of the open system consisting of the matter due to its gravity environment.  But, by (\ref{equalents}), these are actually equal and so, in this environment-paradigm sense, both statements are therefore true, without contradiction.   On the other hand, there is no reason why the third statement should be true in any sense and in fact, on our hypothesis it is clearly not true -- the total entropy being, by (\ref{3entropies}) not the sum of the first two, but rather, equal to each of them. 

The fact that it seems capable of providing this resolution to the thermal atmosphere puzzle provides further support for the validity of our matter-gravity entanglement hypothesis.

\section{The weak string-coupling limit of black-hole equilibrium states and black hole entropy}

Some of the most interesting work towards computing (in certain cases) or, at least, gaining a better understanding of, black hole entropy has been within string theory.  Here I shall briefly recall the basic idea due to Susskind \cite{Susskind} and one particular line of  development by Horowitz and Polchinski \cite{HoroPolch} \cite{Horowitz} which leads to an explanation of how the entropy of spherically symmetric black holes scales with $M^2$ (the square of the black-hole mass), albeit the argument is semi-qualitative and does not tell us the constant term (so doesn't explain the factor of $1/4$ in (\ref{HawkEnt})).

First I will outline the Susskind-Horowitz-Polchinski (SHP) argument.  Then I will criticize it.   Then I will propose a modification of the SHP argument which is free from the criticisms I raise and is consistent with the understanding of black-hole equilibrium states on the matter-gravity entanglement hypothesis that I outlined in Section 3.

\begin{figure}[h]
\centering
\includegraphics[trim = 6cm 22cm 6cm 4cm, clip]{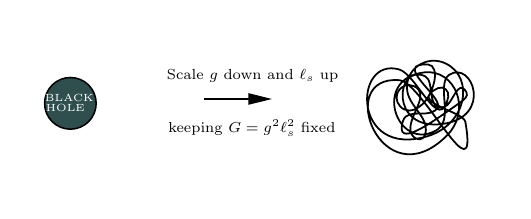}
\caption{The weak string-coupling limit of a black hole is a long string.}
\label{fig:6}      
\end{figure}

The SHP argument \cite{HoroPolch} \cite{Horowitz} is in two steps\footnote{\label{prefactor}We adopt similar simplifications to those adopted in \cite{HoroPolch} \cite{Horowitz}.  Thus the spacetime dimension is taken to be 4 and the power-law prefactors in the densities of states are ignored.   See however \cite{KayMore} for the importance of those prefactors in my proposed modification of the SHP argument.}:  First (see Figure \ref{fig:6}) one argues that, as one scales the string coupling-constant, $g$, down and the string length, $\ell_s$ up, keeping Newton's constant $G=g^2\ell_s^2$ fixed, a black hole goes over to a long string.    This will have density of states (i.e.\ number of states per unit energy, where we use $\epsilon$ to denote energy) 
$\sigma_\mathrm{long string}(\epsilon)$ approximately of the form  of a constant times $e^{\ell_s\epsilon}$.   

Secondly, one equates the entropy, $S_\mathrm{black hole}$, with ``$k\log(\sigma_\mathrm{long string}(\epsilon))$'' $=k\ell_s\epsilon$ at 
$\epsilon=$ constant times $M$ when $\ell_s=$
constant times $GM$ whereupon $S_\mathrm{black hole} =$ constant times 
$kGM^2$.

Our criticism of this is that it is not correct to equate an entropy with the logarithm of a density of states.  (Nor indeed, in other string theory work, with the logarithm of a degeneracy -- see \cite{KayMod} and  \cite{KayEntropy}.)    Indeed it only ever makes sense in physics to take the logarithm of a dimensionless quantity but a density of states has of course the dimensions of inverse energy!   

Our proposed modification of the SHP scenario \cite{KayMod} \cite{KayMore} is to consider, in place of the limit 
\[
\hbox{\textit{black hole}} \rightarrow \hbox{\textit{long string}},
\]
the limit
\[
\hbox{\textit{black hole in equilibrium with thermal atmosphere in a box}} \qquad\rightarrow
\]
\[
\hbox{\textit{long string in equilibrium with atmosphere of small strings in a suitably rescaled box}}.
\]
\begin{figure}[h]
\centering
\includegraphics[trim = 5.5cm 21cm 5.5cm 5cm, clip]{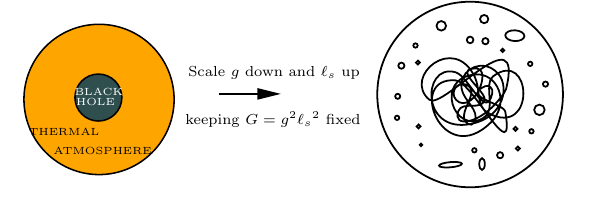}
\caption{The weak string-coupling limit of a black hole in equilibirum with its atmosphere in a suitable box is a long string in equilibrium with its stringy atmosphere in another box.}
\label{fig:7}      
\end{figure}
The key fact \cite{HoroPolch} \cite{Horowitz} about a string equilibrium state of this latter type is that (in a certain approximation where we ignore certain power-law prefactors -- see Footnote \ref{prefactor}) the long string and its stringy atmosphere will have densities of states of the exponential form:
\begin{equation}
\label{stringstatedens}
\sigma_\mathrm{long string}(\epsilon) \sim c e^{\ell_s\epsilon}, \quad \sigma_\mathrm{stringy atmosphere}(\epsilon) \sim c' e^{\ell_s\epsilon}
\end{equation}
where the constants $c$ and $c'$ may be different, but, importantly the exponents are the same.

I have demonstrated (see Section 5 for a discussion of the proof) that:

\medskip

\noindent
{\bf Theorem 1:} \textit{For any pair of weakly coupled systems (to be called here `system' and `bath') with densities of states as in (\ref{stringstatedens}) a randomly chosen {\rm pure} equilibrium state with total energy $E$ will, with very high probability, have a system-bath entanglement entropy approximately equal to $k\ell_s E/4$.   It will also be such that the reduced states of system and bath separately each have energy $E/2$ and are each approximately thermal at  temperature $T=1/k\ell_s$}

\medskip
Applying this theorem and reading `long string' for `system' and `stringy atmosphere` for `bath' (or vice versa) and equating the black hole mass, $M$, with a constant times $E$ and the entanglement entropy of this theorem with the matter-gravity entanglement entropy of  the black hole equilibrium state at $\ell_s=$ constant times $GM$ (as in the unmodified argument) the latter entropy will thus be a constant times $kGM^2$. Thus we achieve a corrected string explanation of this formula for the black hole entropy which is not subject to the criticism we made of the original SHP approach.  Moreoever making the same substitution, $\ell_s=$ constant times $GM$, the temperature formula for the reduced states of the long string and of its stringy atmosphere goes over to the temperature formula
$T=$ a constant times $1/kGM$, which agrees with the Hawking temperature formula (\ref{HawkTemp}) (up to a constant).\footnote{\label{coin} Intriguingly, as pointed out in \cite{KayMod}, if one equates $M$ with $E/2$ and equates $\ell_s$ with $8\pi GM$, then one gets the right value both for the Hawking temperature and the Hawking entropy.   However, as explained in 
\cite{KayMod} and \cite{KayMore} this numerical coincidence should be interpreted with caution.}

That ends my discussion of my matter-gravity entanglement hypothesis and of how it offers a resolution to the three puzzles: the second law puzzle, the black hole information loss puzzle, and the thermal atmosphere puzzle and, finally, in this section, of how it enables a modification of the SHP string approach to black hole entropy which is free from the criticism\footnote{To provide further perspective on that criticism, let us recall that the attempt to provide a microscopic explanation of thermodynamical behaviour in terms of a classical statistical mechanics has often been criticized because it requires the introduction of an ad hoc quantity with the dimensions of action in order to provide a unit of volume in phase space.  It has been said that this shortcoming of classical statistical mechanics is overcome in quantum statistical mechanics where a suitable power of the quantity $\hbar$ effectively provides the right volume element.  One might re-express the main thesis of this section by saying that, in a similar way, the need to introduce an ad hoc dimensionful quantity as in the SHP approach to black hole entropy and the resolution of that difficulty along the lines explained in the main text indicates that, to have a satisfactory microscopic explanation of thermodynamical behaviour, a quantum statistical mechanics is \textit{also} insufficient and what is needed, instead, is a quantum-gravitational statistical mechanics based on our matter-gravity entanglement hypothesis.} which I made of the original SHP approach.

In the remainder of the talk I would like to supply some of the details about how I proved the above theorem.

\section{Explanations of thermality: traditional and modern}

Theorem 1 in fact relies on a general theorem -- which is stated below as Theorem 2 -- which I obtained \cite{KayThermality} in a general setting where one has a total system (in \cite{KayThermality} I abbreviate this with the the term `totem' and I shall follow that terminology here) consisting of a (quantum) system weakly coupled to an energy bath.

Such a totem will have a Hamiltonian of form
\[
H=H_\mathrm{system} + H_\mathrm{bath} + H_\mathrm{interaction} 
\]
on
\[
{\cal H}_\mathrm{system}\otimes {\cal H}_\mathrm{bath}
\]
where $H_\mathrm{interaction}$ is assumed to be sufficiently weak that it can be ignored for the purposes of counting energy levels;
${\cal H}_\mathrm{system}$ and ${\cal H}_\mathrm{bath}$ each have positively supported, locally finite, discrete spectrum with monotonically increasing densities of states,
\[
\sigma_\mathrm{system}(\epsilon) \quad\hbox{and}\quad \sigma_\mathrm{bath}(\epsilon).
\]

Theorem 2 may be considered to generalize a result of Goldstein, Lebowitz, Tumulka and Zanghi (GLTZ) \cite{GoldsteinEtAl} (see also \cite{PopescuEtAl}) which explains why it is that a small system in contact with a large energy bath will typically be in an (approximate) thermal equilibrium state.  So I will first briefly recall that result:

\subsection{Thermality in the case the system is small}

The GLTZ explanation is itself a modern replacement for the earlier traditional explanation of the thermality of a small system in contact with a heat bath, so let me recall that first.

\begin{figure}[h]
\centering
\includegraphics[trim = 5.5cm 21cm 5.5cm 5cm, clip]{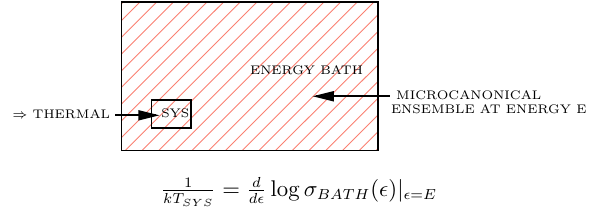}
\caption{The traditional explanation of the thermality of a small system}
\label{fig:8}      
\end{figure}

The traditional explanation is based on a mathematical theorem which tells us that if the totem is in a microcanonical ensemble with energy in a narrow band around some total energy $E$, then the small system will be approximately in a thermal equilibrium state with temperature, $T_\mathrm{system}$ given by the formula (note that the dimensionful argument of the logarithm is innocuous here because the logarithm is differentiated):
\begin{equation}
\label{trad}
{1 \over kT_\mathrm{system}}={d\over d\epsilon} \log\sigma_\mathrm{bath}(\epsilon)|_{\epsilon=E}.
\end{equation}
The modern explanation \cite{GoldsteinEtAl} is based on a mathematical theorem (proven in \cite{GoldsteinEtAl}) that if the totem state is a pure state, randomly chosen from the set of all pure states with totem energy in a narrow band around $E$ (where the random choice is with respect to a natural measure on the set of all  these pure states) then the small system will very probably be very close to the same thermal equilibrium state with a temperature given by the same formula (\ref{trad}).

\begin{figure}[h]
\centering
\includegraphics[scale=0.8, trim = 4.5cm 21cm 4cm 5cm, clip]{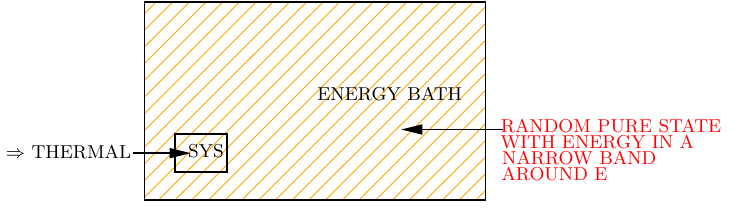}
\caption{The modern explanation of the thermality of a small system}
\label{fig:9}      
\end{figure}

The advantage of the ``modern'' over the ``traditional'' point of view is that it bases a theory of how systems get themselves into (approximate) Gibbs states on the same foundational assumption that we usually make for the foundations of quantum mechanics -- namely that the total state of a full closed system is a pure (vector) state.

\subsection{What happens when System and Energy Bath are of comparable size?}

One might think that one could apply the GLTZ result directly to the case our totem is the string equilibrium state illustrated in Figure \ref{fig:7}, identifying, say, the long string with our `system' and the stringy atmosphere with our `energy bath'.  However, neither of these can be regarded as small with respect to the other.  Here we should clarify that `small' in this context would mean having much more widely spaced energy levels, i.e.\ having a much lower density of states.  Instead both densities of states are (ignoring the power-law prefactors I mentioned earlier) of the exponentially increasing form (\ref{stringstatedens}).

It turns out in general, that when the system and the energy bath are of comparable size, then -- on both the traditional assumption of a totem microcanonical ensemble and the modern assumption of a random total pure state with energy in a small band --  it is no longer necessarily the case that either system or energy bath will probably be in a thermal equilibrium state.  However, I have shown \cite{KayThermality} with regard to the modern approach: 

\medskip

\noindent
{\bf Theorem 2}  There is a special density operator (see the Appendix for details)
\begin{equation}
\label{special}
\rho^\mathrm{modapprox}_\mathrm{system} \ \ \hbox{on} \ \ {\cal H}_\mathrm{system}
\end{equation}
such that, given a random vector, $\Psi\in {\cal H}_\mathrm{system}\otimes {\cal H}_\mathrm{bath}$, with energy in a narrow band around $E$, then the partial trace of $|\Psi\rangle\langle\Psi|$ over ${\cal H}_\mathrm{bath}$ is very probably very close to $\rho^\mathrm{modapprox}_\mathrm{system}$.

\smallskip

\noindent
(And similarly with \textit{system} $\leftrightarrow$ \textit{energy bath}).

\bigskip

But it is important to realize that when system and energy bath are \textit{of comparable size}, $\rho^\mathrm{modapprox}_\mathrm{system}$ is \textit{not always thermal}.  (And neither, by the way, is the reduced state of the system thermal when the total state is in a traditional microcanonical ensemble.)

\begin{figure}[h]
\centering
\includegraphics[width = 200pt, height = 200pt]{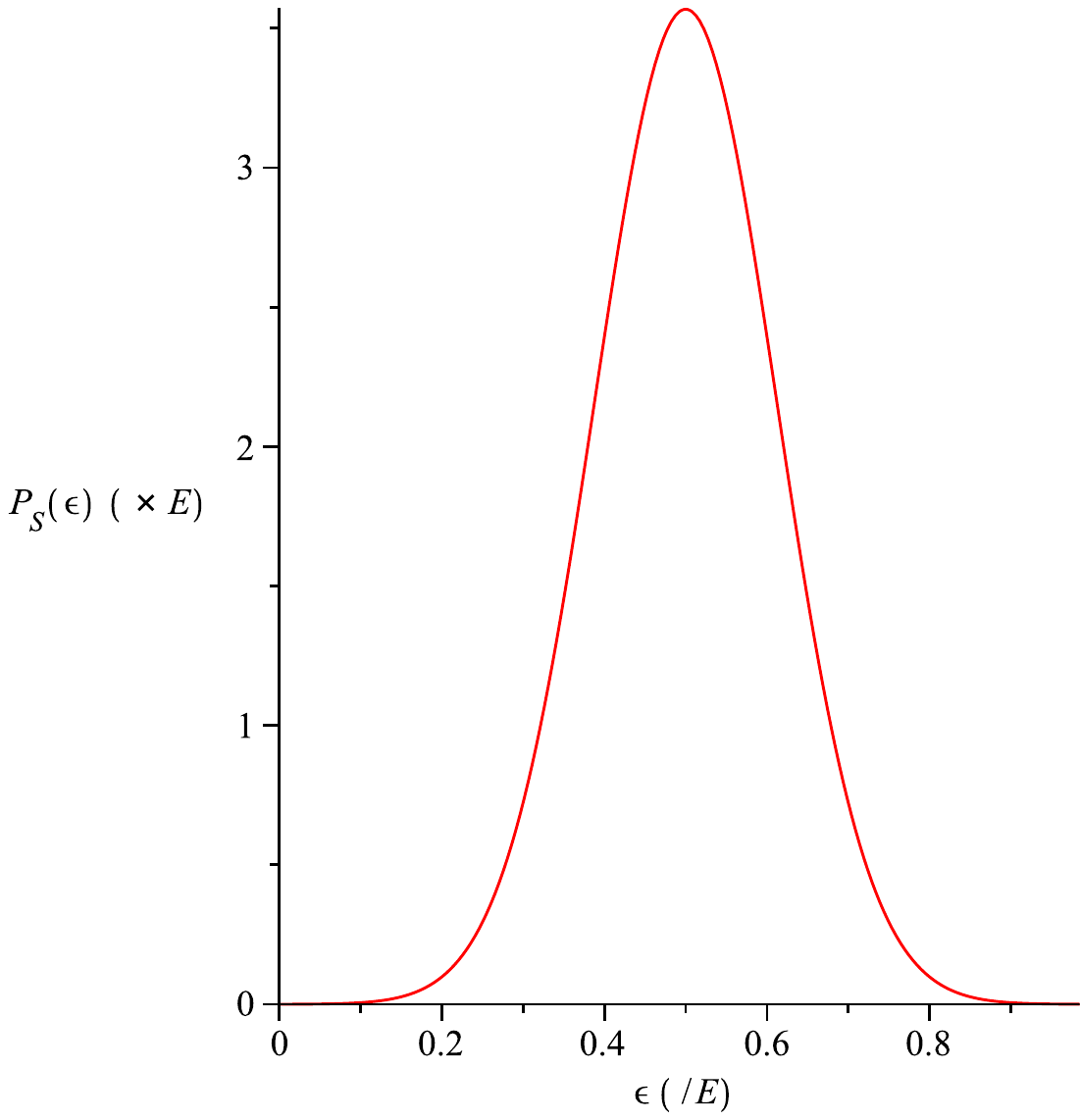}
\caption{Plot of the energy probability density, $P_\mathrm{system}(\epsilon)$,
when \textit{system} and \textit{energy bath} have the same power-law density of states
$\sigma(\epsilon)=A\epsilon^{N}$ for the (`unusually' small) value $N=10$}
\label{fig:10}
\end{figure}

\begin{figure}[ht]
\centering
\includegraphics[width = 200pt, height = 200pt]{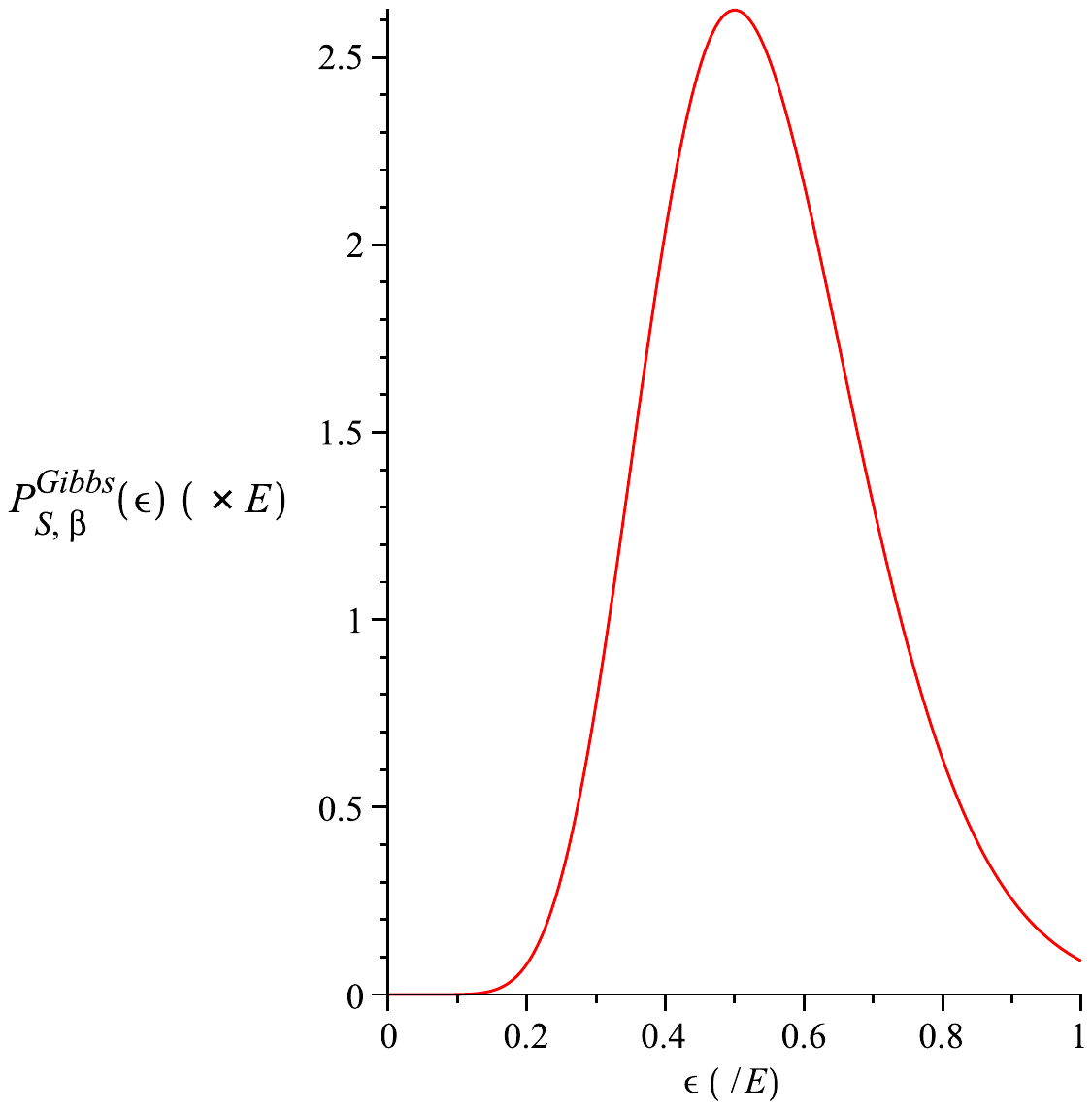}
\caption{Plot of the energy probability density, $P^{\mathrm{Gibbs}}_\mathrm{system,\beta}(\epsilon)$ for the thermal state at inverse temperature, $\beta$, on our \textit{system} with density of states 
$\sigma(\epsilon)=A\epsilon^N$, for the same (`unusually' small) value $N=10$ and for $\beta=22/E$ (i.e.\ the value of $\beta$ for which the mean energy is $E/2$).}
\label{fig:11}
\end{figure}

E.g.\ if $\sigma_\mathrm{system}(\epsilon)$ and $\sigma_\mathrm{bath}(\epsilon)$ take, respectively, the power law forms $\sigma_\mathrm{system}(\epsilon)=A_S\epsilon^{N_S}$, $\sigma_\mathrm{bath}(\epsilon)=A_S\epsilon^{N_S}$  (the typical behaviour of ordinary matter when $N_A$ and $N_B$ are comparable in size to Avogadro's number) then the system `energy probability density', $P_\mathrm{system}(\epsilon)$ \cite{KayThermality} will be a Gaussian (in fact the same Gaussian on both traditional and modern assumptions) rather than the Gibbsian distribution characteristic of a thermal state.  See Figures \ref{fig:10} and \ref{fig:11}.

\subsection{The special nature of exponential densities of states}

However, it is shown in \cite{KayThermality}, regarding the modern approach\footnote{A similar result to Theorem 3 holds for the traditional (microcanonical) approach, except that (now neglecting logarithmic terms) in place of $k\ell_s E/4$ one finds \cite{KayThermality} that the system and the energy bath have entropy $k\ell_s E/2$.   The difference between these two results is interesting since it demonstrates that, in general, the traditional and modern approaches don't give the same results.  (It is also interesting since the ``right value for the Hawking entropy'' mentioned in Footnote \ref{coin} depends on the denominator being 4 -- rather than 2).}

\smallskip
\noindent
{\bf Theorem 3:}

When system and energy-bath densities of states both take the exponential
form of Equation (\ref{stringstatedens}):

\smallskip

\begin{itemize}

\item $\rho^\mathrm{modapprox}_\mathrm{system}$ and $\rho^\mathrm{modapprox}_\mathrm{bath}$ are (close to\footnote{See \cite{KayThermality} for the sense in which these states are close to thermal.}) \textit{thermal} at  temperature $T=1/k\ell_s$. (And each have mean energy $E/2$.)

\smallskip

\item Also, the \textit{system-energy bath entanglement entropy}, $S$,
$\left(=S^\mathrm{vN}(\rho^\mathrm{modapprox}_\mathrm{system})=S^\mathrm{vN}(\rho^\mathrm{modapprox}_\mathrm{bath})\right)$ is approximately $k\ell_s E/4$.
\footnote{The exact result \cite[Endnote 29]{KayThermality} is $k\ell_s E/4 + k\log(c_\mathrm{S}c_\mathrm{B}E^2)/2 - k(\log(c_\mathrm{S}/c_\mathrm{B}))^2/4E$.}

\end{itemize}

Theorem 1 of Section 4 clearly follows immediately from Theorems 2 and 3.

\section*{Appendix: Details on ${\rho^\mathrm{modapprox}_\mathrm{system}}$}

In this appendix we give the detailed formula for the special density operator (\ref{special}). 

Define the ($M$-dimensional) Hilbert space, ${\cal H}_M$ ($M$ assumed large) consisting of elements with total energy in a narrow band $[E, E+\Delta]$ 
to be the closed span of eigenstates of the total Hamiltonian with energies, $\epsilon\in [E, E+\Delta]$.

For convenience, replace the system of interest by a system with equally spaced energy levels with spacing equal to $\Delta$ -- each energy level, $\epsilon$, having 
degeneracy, $n(\epsilon)=\sigma(\epsilon)\Delta$ (so that the new system will have the same density of states, $\sigma(\epsilon)$ as the original system).

\smallbreak

\noindent
(Note that then $M=\sum_{\epsilon=\Delta}^E n_\mathrm{system}(\epsilon) n_\mathrm{bath}(E-\epsilon)$.)

\smallbreak

We note first that the traditional microcanonical density operator, $\sum_\mathrm{basis\,for\,{\cal H}_M} |\psi_i\rangle\langle\psi_i|$ is then easily seen to have reduced density operator on ${\cal H}_\mathrm{system}$ equal to
\[
\rho^{\mathrm{microc}}_{\mathrm{system}}=M^{-1}\sum_{\epsilon=\Delta}^E
n_{\mathrm{bath}}(E-\epsilon) \sum_{i=1}^{n_{\mathrm{system}}(\epsilon)}
|\epsilon, i\rangle\langle \epsilon, i|
\]
where $|\epsilon, i\rangle$ denotes a basis for the $n_\mathrm{system}(\epsilon)$-dimensional degeneracy subspace of ${\cal H}_\mathrm{system}$ with energy $\epsilon$ (assumed to be a multiple of $\Delta$) and the sum over $\epsilon$ is over multiples of $\Delta$.

The \textit{modern} replacement for this result is that a random pure density operator, 
$|\Psi\rangle\langle\Psi|$, on ${\cal H}_M$ will have a 
reduced density operator on ${\cal H}_\mathrm{system}$ which (as is argued in \cite{KayThermality}) is very probably very close to ${\rho^\mathrm{modapprox}_\mathrm{system}}$ where
\[
\rho^{\mathrm{modapprox}}_{\mathrm{system}}=M^{-1} \ \ \hbox{times}
\]
\[
\sum_{\epsilon=\Delta}^{E_c}
n_{\mathrm{bath}}(E-\epsilon)\sum_{i=1}^{n_{\mathrm{system}}(\epsilon)}
 |\epsilon, i\rangle\langle \epsilon, i|+\sum_{\epsilon=E_c+\Delta}^E
n_{\mathrm{system}}(\epsilon)\sum_{i=1}^{n_{\mathrm{bath}}(E-\epsilon)}
|\widetilde{\epsilon, i}\rangle\langle
\widetilde{\epsilon, i}| 
\]
where $E_c$ is the energy at which $\sigma_\mathrm{system}(\epsilon)=\sigma_\mathrm{bath}(E-\epsilon)$ and the $|\widetilde{\epsilon, i}\rangle$ span
an orthonormal basis of an $n_\mathrm{Bath}(E - \epsilon)$-dimensional subspace of the $n_\mathrm{System}(\epsilon)$-dimensional) energy-$\epsilon$
subspace of ${\cal H}_\mathrm{System}$ which depends on $\Psi$ in a way explained in detail in  \cite{KayThermality}.

\section*{Afterword}

To end, let me mention some related aspects of the matter-gravity entanglement hypothesis that we have not had time to discuss.    One is an extension of the theory beyond closed systems to include open systems.   For this, we refer to \cite[Endnote (xii)]{KayAbyaneh} or \cite{KayEntropy}.   Another concerns the relevance of the matter-gravity entanglement hypothesis to the measurement problem in quantum mechanics and a possible resolution to the Schr\"odinger Cat puzzle.  For this, see \cite{Kay1998a}, \cite{Kay1998b} and \cite{KayAbyaneh}.  Finally, the papers \cite{KayInstab}, \cite{KayLupo} (see also \cite{KayEntropy} for a brief outline of this work) include a discussion of a possible mechanism whereby, when one passes from a quantum field theory in curved spacetime description to a description in which the backreaction of the stress-energy tensor on the metric is taken into account, the horizon of an enclosed (say Kruskal) black hole becomes unstable with the consequence that entanglement between the right and left Kruskal wedges in a quantum theory in curved spacetime context transmutes into entanglement between matter and gravity -- in support of the solution to the thermal atmosphere puzzle presented in Section 3.

\end{document}